\documentclass[12pt, a4paper]{scrartcl}
\usepackage[utf8x]{inputenc}
\usepackage{graphicx}
\usepackage{amsmath,amsfonts,amssymb}
\usepackage{amsxtra,amsthm}
\usepackage{amstext}
\usepackage{hyperref}
\usepackage{booktabs}
\usepackage[numbers,square,sort&compress]{natbib}
\usepackage{type1cm}
\usepackage{eso-pic}
\usepackage{color}
\usepackage{array}
\usepackage[noblocks]{authblk}
\usepackage{lscape}

\usepackage{lineno}

\addtokomafont{caption}{\small}
\setkomafont{captionlabel}{\sffamily\bfseries}
\setcapindent{1em}

\newcommand*{\rom}[1]{\expandafter\@slowromancap\romannumeral #1@}

\begin{document}

\title{{The effect of interference on the $\mathrm{CD}8^+$ T cell escape rates in HIV}}
\author[1,*]{Victor Garcia}
\author[1]{Roland R. Regoes}
\affil[1]{{\footnotesize Institute of Integrative Biology, ETH Zurich, Universit\"atstr. 16, CH-8092 Zurich, Switzerland}}
\affil[*]{{\footnotesize \emph{corresponding author:} P: +41-44-633 60 34, email: victor.garcia@env.ethz.ch}}
\date{}

\maketitle

\section*{Abstract}
In early HIV infection, the virus population escapes from multiple $\mathrm{CD}8^+$ cell responses. 
The later an escape mutation emerges, the slower it outgrows its competition, i. e. the escape rate is lower. This pattern could indicate that the strength of the $\mathrm{CD}8^+$ cell responses is waning, or that later viral escape mutants carry a larger fitness cost.
In this paper, we investigate whether the pattern of decreasing escape rate could also be caused by genetic interference among different escape strains. 
To this end, we developed a mathematical multi-epitope model of HIV dynamics, which incorporates stochastic effects, recombination and mutation. We used cumulative linkage disequilibrium measures to quantify the amount of interference.
We found that nearly synchronous, similarly strong immune responses in two-locus systems enhance the generation of genetic interference. This effect, combined with densely spaced sampling times at the beginning of infection, leads to decreasing successive escape rate estimates, even when there were no selection differences among alleles. These predictions are supported by experimental data from one HIV-infected patient.
Thus, interference could explain why later escapes are slower. Considering escape mutations in isolation, neglecting their genetic linkage, conceals the underlying haplotype dynamics  
and can affect the estimation of the selective pressure exerted by CD$8^+$ cells. In systems in which multiple escape mutations appear, the occurrence of interference dynamics should be assessed by measuring the linkage between different escape mutations.


\section*{Introduction}

CD$8^+$ T cell responses exert strong selection pressures on Human Immunodeficiency Virus (HIV) as shown by evidence stemming from a variety of observations  \cite{goulder2004hiv, schmitz1999control, friedrich2004reversion, barouch2005dynamic, kent2005reversion, peut2006fitness, crawford2007compensatory, frater2007effective, li2007rapid}. The selective pressure exerted by 
these responses has been quantified from the growth rate advantage of viral mutants that escape CD$8^+$ T cell control \cite{fernandez2005rapid,asquith2006inefficient,mandl2007estimating}.

More recently, studies based on the analysis of the entire viral genome revealed escape in multiple epitopes targeted by CD$8^+$ T cell responses \cite{turnbull2009kinetics, goonetilleke2009first}. 
A mathematical analysis of these data shows that late-emerging escape mutants outgrow the resident virus population more slowly than early escapes \cite{asquith2006inefficient, goonetilleke2009first,ganusov2011fitness,henn2012plospath,althaus2008dynamics}
We refer to this pattern as \emph{escape rate decrease} (ERD).

ERD has been assumed to have a biological basis, arising from either variation in \cite{ganusov2011fitness} or a decrease of fitness advantages of escape mutants during the course of infection \cite{ganusov2011fitness, ganusov2013jsm, althaus2008dynamics}. But it could also be due to complex dynamical interactions between escape mutations in different epitopes. The complexity arises from the fact that two similarly beneficial mutations rarely arise simultaneously on one genome. Rather, they arise on different genomes which, after outcompeting the wild type strain, enter a state of competition. 
This state is eventually resolved by one mutation going to fixation by chance, or by one of the strains acquiring the other beneficial mutation. In this scenario, the fixation of one of the advantageous mutations takes longer than its fitness advantage predicts. In population genetics, this complex dynamics is called \emph{interference} \cite{hill1966effect, gerrish1998fate, desai2007genetics}. 

\begin{figure}[h]
\begin{center}
\includegraphics[scale = 0.4]{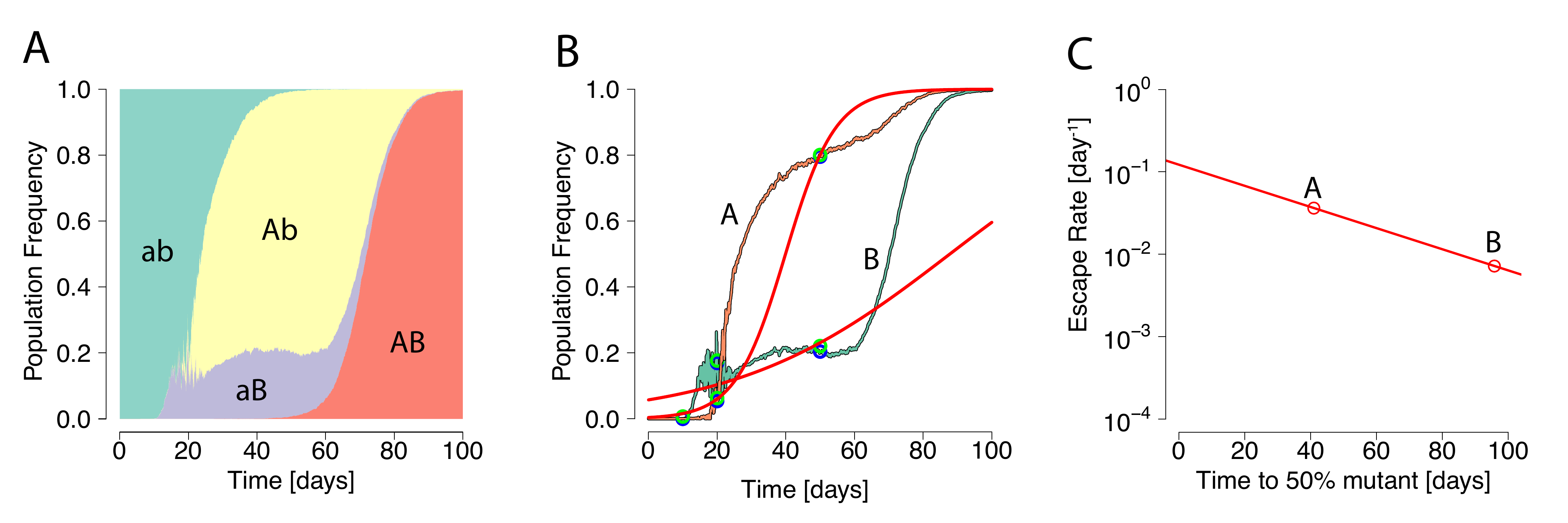} 
\end{center}
 \caption{\textbf{Pattern of ERD emerging from repeated application of logistic model fits in an interference scenario.} (A) Population frequencies of two-locus system haplotypes display interference. The wildtype ab (green area) gives rise to two beneficial single mutants Ab and aB (yellow and violet areas, respectively). (B) Fixation patterns of the beneficial alleles A (orange line) and B (green line). Samples of frequencies of A and B are taken at times 10, 20 and 50 days (blue points). Logistic model fits (red lines) are laid through sample points with an added noise (green points). (C) From each logistic model fit the escape time and escape rate are calculated. A pattern of ERD is generated due to interference. }
 \label{fig:selective_interference}
\end{figure}

Figure \ref{fig:selective_interference} shows how interference can affect the interpretation of ERD in early HIV infection. Describing the escape dynamics with models that reduce the dynamics to the competition between two types only --- a wildtype and a single escape mutation --- and neglect interference, misrepresents the selection pressures at work, and may lead to biased estimates of their strength. 

In this study, we investigated under which circumstances genetic interference arises and may thus lead to misinterpretation of pattern of escape rate changes in systems with two loci. 
To this end, we developed a virus dynamics model, in which viruses possess multiple epitopes and can escape from CD$8^+$ T cells directed against them. The model builds on well-established work \cite{nowak2000virus,perelson2002modelling,asquith2006inefficient,deBoer2007jv,althaus2008dynamics,fryer2010modelling}, and is stochastic to describe mutation, extinction and fixation of virus strains adequately, and allows recombination of viruses. 

We found that interference emerges mainly when CD$8^+$ T cell responses coincide and are similarly strong, but only in systems with a high level of stochasticity. This interference leads to ERD if the virus population was sampled more often early than late, a scheme commonly adopted in empirical studies \cite{goonetilleke2009first}.
We tested these predictions in early-infection data from an HIV-positive patient obtained by Henn et al.\ \cite{henn2012plospath} and subsequently reconstructed to haplotypes \cite{pandit2013arxiv}.

\section*{Materials and Methods}

Here, we extend the model of Althaus and De Boer \cite{althaus2008dynamics}, which is in turn based on earlier work \cite{nowak2000virus,perelson2002modelling,asquith2006inefficient,deBoer2007jv,
fryer2010modelling}. 
In our model, a viral strain $\textbf{i}$ is assumed to present $n$ different viral epitopes to the hosts' immune system. The strain is represented by a string of binary digits, where a 1 at the $j^{\mathrm{th}}$ entry signifies the presence of an escape mutation in the $j^{\mathrm{th}}$ epitope. A 0 at the same entry signifies no mutation, and the strains is recognized by epitope-specific immune responses $E_j(t)$. 
The model equations are:

\begin{align}
\label{eq:dynamics}
\frac{d}{dt}T = &\sigma - d_TT - \sum_\textbf{i }\frac{\beta TpP_\textbf{i}}{h_\beta + T}\\
\frac{d}{dt}I_\textbf{i} = &\frac{\beta TpP_\textbf{i}}{h_\beta + T} -dI_\textbf{i} -\gamma I_\textbf{i} 
+\sum_\textbf{x} \left( m_{\textbf{x}\textbf{i}} I_\textbf{x} - m_{\textbf{i}\textbf{x}} I_\textbf{i}  \right) \nonumber  \\
& + r \cdot t\cdot I_{tot}
\left( \sum_{\textbf{x,y} \in Q(\textbf{i})}\frac{P_\textbf{x} P_\textbf{y}}{P_{\mathrm{tot}}^2}\frac{1}{2^{d(\textbf{x,y})}} \omega'(\textbf{x,y,i}) 
 - \sum_\textbf{x}  \frac{2P_\textbf{x} P_\textbf{i}}{P_{\mathrm{tot}}^2}\frac{2^{d(\textbf{x,i})}-1}{2^{d(\textbf{x,i})}}(-\omega'(\textbf{x,i,i})) \right) \\
\frac{d}{dt}P_\textbf{i} = & \gamma I_\textbf{i} - \delta P_\textbf{i} - k \sum_j^n \left( \frac{ a_{j\textbf{i}} E_j }{ h_k + \sum_\textbf{x} a_{j\textbf{x}}P_\textbf{x} + \sum_s^n a_{s\textbf{i}} E_s }\right) P_\textbf{i}.
\end{align}

\subsection*{Target cells}
We assume a compartment of $\mathrm{CD}4^+$ target cells $T$, which is replenished at rate $\sigma$ and gets naturally depleted at a rate $d_T$ per cell. Virions of type $\textbf{i}$, $V_\textbf{i}$,  will infect target cells at a rate $\frac{\beta TV_\textbf{i}}{h_\beta + T}$ and produce infected cells ($I_i$), where $\beta$ is the maximum infection rate per day for a virus particle, and $h_\beta$ is the target cell density where the infection rate is half-maximal \cite{deBoer2007jv,althaus2008dynamics}. Viral load and productively infected cells of type $\textbf{i}$ ($P_\textbf{i}$) are connected by $\frac{dV_i}{dt} = fP_i - d_VV_i$. We assume that $V_\textbf{i}$ and $P_\textbf{i}$ are fast coupling. Hence, $V_\textbf{i} = p P_\textbf{i}$, where $p = \frac{f}{d_V}$ is the net production rate per cell. 

\subsection*{Infected cells}
Target cells infected with strain $\textbf{i}$  ($I_\textbf{i}$) die at rate $d$ and enter an eclipse phase at rate $\gamma$, after which they become productively infected. 

\subsubsection*{Mutation}
Cells infected with strains $\textbf{x}$ are converted to cells infected with strain $\textbf{y}$ at rate $m_{\textbf{xy}}$. $m_{\textbf{xy}}$ is the locus-wise product of the probabilities for an epitope in strain $\textbf{x}$ to be mutated into the corresponding epitope in strain $\textbf{y}$. We distinguish between forward mutations and reverse mutations. Forward mutations change a 0 allele into a 1. Reverse mutations do the opposite. An epitope is assumed to consist of about $m = 8$ codons. The mutation rate is $3 \cdot 10^{-5}$ per bp per replication \cite{mansky1995jv}. For an escape epitope to emerge, this amounts to a rate of $\approx 1 \times 10^{-4}$ per epitope per replication. The reverse mutation rate was set to $\approx 5 \times 10^{-7}$ per epitope per replication (see \textit{Electronic Supplementary Material (ESM)}).

\subsubsection*{Recombination}

Cells infected with strain $\textbf{i}$ can arise and be lost by recombination. Both processes are assumed to occur at the same baseline recombination rate $r$ and  to be proportional to the fraction of co-infected cells in the population $t \approx 5 \cdot 10^{-3}$, \cite{jung2002nature, josefsson2010ath, neher2010pcb, batorsky2011pnas}. We chose the baseline recombination rate $r$ to incorporate those rates which do not directly depend on the strain  frequencies or types. 

The terms within brackets in the recombination term in equation 
(2) deal with probabilities that depend on the strain frequencies. The first sum encompasses recombination events that increase $I_\textbf{i}$. All pairs $(\textbf{x,y})$ that can recombine into $\textbf{i}$, $Q(\textbf{i})$, are considered. Each such pair $(\textbf{x,y})$ coinfected a cell with probability $\frac{P_\textbf{x} P_\textbf{y}}{P_{\mathrm{tot}}^2}$, where  $P_{\mathrm{tot}} = \sum_\textbf{j}P_\textbf{j}$. 
The probability of $\textbf{i}$ to be the recombinant offspring is $1/2^{d(\textbf{x,y})}$, where $d(\textbf{x,y})$ denotes the Hamming distance between two strains.

Infected cell numbers should remain unaltered by the action recombination, since recombination only reshuffles alleles. We account for this by using the weights $\omega '(\textbf{x,y,i})$. 
We defined $\omega '(\textbf{x,y,i}) = 2 - 1_{\textbf{x}=\textbf{i}} - 1_{\textbf{y} = \textbf{i}}$, where $1_{\textbf{y} = \textbf{i}}$ is one if $\textbf{y} = \textbf{i}$ and zero otherwise (see \textit{ESM} for details on $\omega '$). 

The second type of events that can decrease $I_\textbf{i}$ are those in which the strain $\textbf{i}$ can recombine with any other into a strain different from $\textbf{i}$. 
Pairs can be formed with all other strains, including itself with probability $\frac{2P_\textbf{x} P_\textbf{i}}{P_{\mathrm{tot}}^2}$. Such an event needs to be weighted with the probability that $\textbf{i}$ will yield offspring distinct to itself: $\frac{2^{d(\textbf{x,i})}-1}{2^{d(\textbf{x,i})}}$. Again, in order to keep infected cell numbers unaffected by recombination, the weight $-\omega'(\textbf{x,i,i})$ is factored in. 


The baseline recombination rate $r$ incorporates several rates and probabilities. These include the probability for co-packaging two parent strains correctly, the number of newly infected cells produced by a single infected cell, the template switching rate and an assumption about the average distance between escape mutations on the genome. We set $r \approx 1.4 \cdot 10^{-4} \cdot \mathrm{day}^{-1}$ for our simulations (see Fig.\ \ref{fig:recombination}).

\subsection*{Productively Infected Cells}
Productively infected cells infected with a strain $\textbf{i}$ are generated at rate $\gamma$ from the eclipsed population $I_\textbf{i}$, and die with rate $\delta$. The CD$8^+$ T cells specific to the epitope $j$, $E_j$, clear productively infected cells whose epitopes they recognize maximally at rate $k$, with $h_k$ the Michaelis-Menten constants \cite{althaus2008dynamics}. The coefficients $a_{j\textbf{i}}$ are 1 if $\textbf{i}$ has a zero at its $j^{\mathrm{th}}$ position, and zero otherwise. They thereby encode the recognition of non-escape epitopes by the CD$8^+$ T cells.

\subsection*{CD$8^+$ T cells}


Unlike previous models describing CD$8^+$ T cell escape  \cite{deBoer2007jv,althaus2008dynamics}, our model does not dynamically link the immune response to the level of viral antigen. In our model, $E_j(t)$ is a numerically constructed fixed time course of the CD$8^+$ T cells of type $j$, consistent with a program-type dynamics established for CD$8^+$ T cell responses against infections in mice \cite{murali1998immunity, ahmed1996science, antia2003jtb, kaech2002natrevimm, kaech2002cell, antia2005natrevimm}.

In order to attain a descriptive CD$8^+$ T cell function we combined two exponential functions and a constant value. The default CD$8^+$ T cell-function starts growing from a single cell at time $t=0$ at a rate of $r_c = 0.9$ per day \cite{davenport2004jv, deBoer2007jv}, until it reaches an upper limit $C$, which is 1.5 logs larger than the final level value $K$: $C = 10^{\log_{10}(K) + 1.5}$ \cite{murali1998immunity}. Once $C$ has been reached, the function declines exponentially at a rate of $0.4 \cdot r_c$, until it reaches $K$. For larger time values, it remains at $K$. In this study, we restricted the \textit{final levels} $K$ to values of the order $10^7$, in accordance with \cite{murali1998immunity, ahmed1996science}. 

\subsection*{Implementation}
For all simulations described in this paper, we used parameters as shown in table \ref{table:PAR}. To implement the dynamics we used the R language for statistical computing \cite{Rmanual} and \textit{adaptivetau} \cite{adaptivetau} for the simulation of \eqref{eq:dynamics} by Gillespie algorithm. The rates of the Gillespie algorithm are the terms on the right-hand side of the ordinary differential equations \eqref{eq:dynamics}.


\begin{table}[h]

\caption{
\bf{Parameter values employed for simulations}}
\begin{tabular}{llr}
\hline
Parameter & Description & Value  \\
\hline
$\sigma$  & replenishment rate of target cells $T$ & $10^8$   $\frac{\mathrm{cells}}{\mathrm{day}}$ \\
$d_T$ & natural rate of target cell death & $10^{-2}$  $\mathrm{day}^{-1}$ \\
$p$   & net virion production rate per productively infected cell  & $10^4$  $\frac{\mathrm{virions}}{\mathrm{day}}$ \\
$\beta$   & maximum infection rate per day for a virus particle  & $5.5 \cdot 10^{-4}$   $\frac{1}{\mathrm{day}}$ \\
$h_{\beta}$ & target cell number at which infection rate is half-maximal & $5 \cdot 10^7$    $\mathrm{cells}$ \\
$d$ & natural rate of infected cell death & $2\cdot 10^{-2}$   $\frac{1}{\mathrm{day}}$ \\
$\gamma$ & transition rate to productively infected cells (eclipse phase) & 1   $\frac{1}{\mathrm{day}}$ \\
$\delta$ & natural rate of productively infected cell death & 1  $\frac{1}{\mathrm{day}}$  \\
$k$ & maximum killing efficiency  &  $50$   $\frac{1}{\mathrm{day}}$ \\
$h_k$ &  cell number at which killing rate is half-maximal &  $10^9$   $\mathrm{cells}$ \\
$\mu$ & viral mutation rate &  $3\cdot 10^{-5}$   $\frac{1}{\mathrm{bp}\cdot \mathrm{replication}}$ \\
$r$ & base recombination rate per replication &  $0.25 \cdot \beta $   $\frac{1}{\mathrm{day}}$ \\
\hline
\end{tabular}


\label{table:PAR}
\end{table}

\subsection*{Stochasticity induced by system rescaling}
HIV's large census population size does not imply a small role of stochastic effects in its dynamics \cite{kouyos2006stochastic}. Hence, we implemented methods to rescale the model system to match a population size of $10^6$ (see \textit{ESM}). By downsizing or magnifying a system we refer to transforming the system under consideration $S$, with its variables $T, I_i, P_i$ into a system downsized by a factor $a$, $S_a$, with corresponding variables $T_a, I_{i,a}, P_{i,a}$. Under a deterministic framework, the ratios between time courses of the downsized variables and the corresponding original system variables are $1/a$.

\section*{Results}

\subsection*{Virus dynamics model reproduces escape dynamics}
\begin{figure}
\begin{center}
\includegraphics[scale = 0.45]{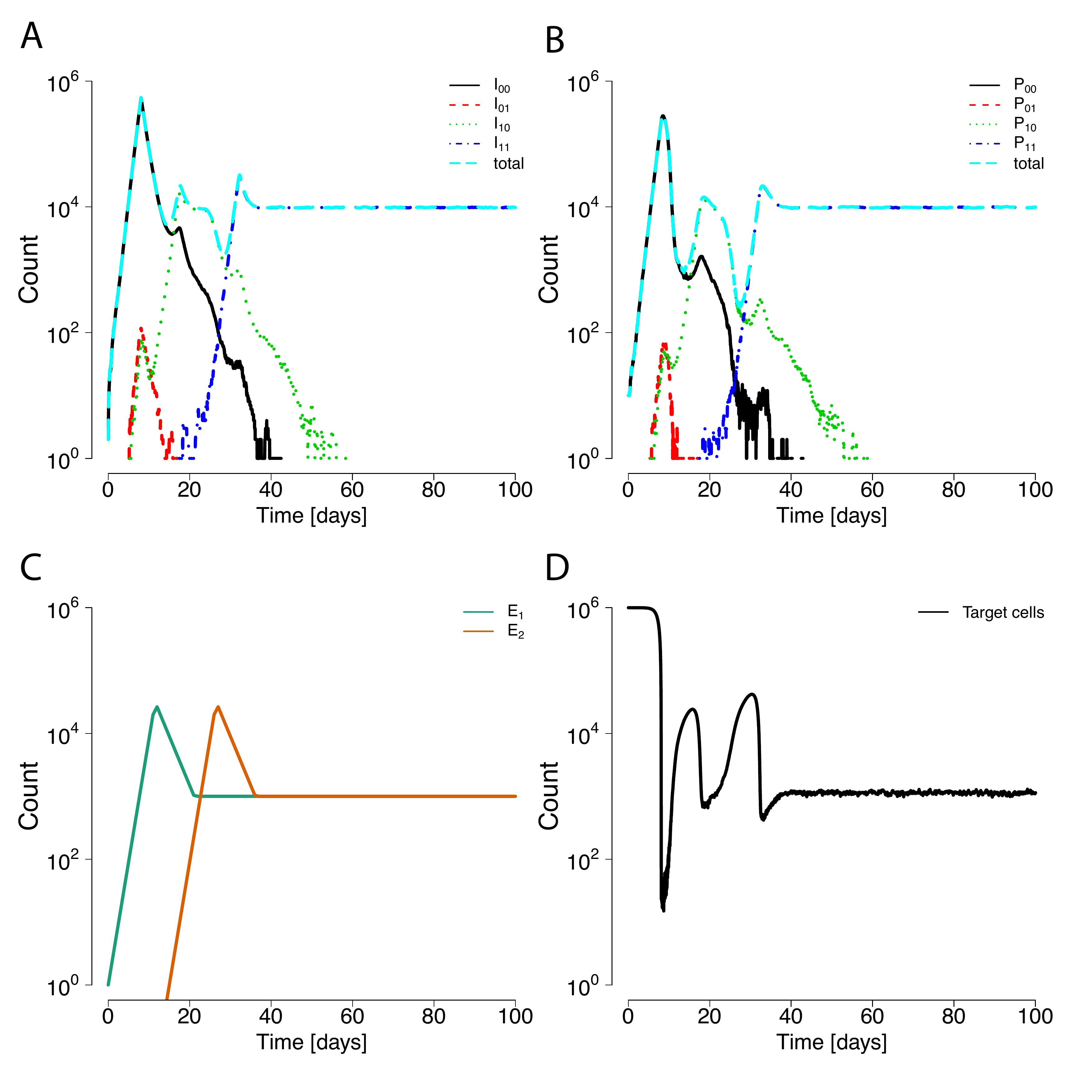} 
\end{center} 
 \caption{\textbf{Example for a simulation run showing sequential escapes in a scaled down two-locus two-allele system.} A) Time course of the number of non-productively infected cells by strain types show sequential transitions from the wildtype to a single escape mutant to a double mutant. 
B) Analogous situation for the time courses of productively infected cells. Productively infected cells are cleared by epitope-specific CD$8^+$ T cell action. Immune response mounting leads to a transitory decrease of the total number of productively infected cells. 
C) The CD$8^+$ functions $E_1$ and $E_2$ start at 0 and 15 days, respectively, with a settling value of $10^{7}$. The mounting of the immune response coincides with the temporary reduction of infected and productively infected cell numbers. D) The killing of productively infected cells causes the transitory reduction of produced virions, temporarily reducing net new infections and releasing target cells. Parameters are as given in table \ref{table:PAR}.}
 \label{fig:time_course}
\end{figure}

The model presented here reproduces basic experimentally observed aspects of HIV/SIV dynamics (Fig.\ref{fig:time_course}). In our model, the initial growth of the virus has been gauged to $1.2 \pm 0.1$ per day in accordance with \cite{ribeiro1998frequency}. Consistent with clinical and experimental data \cite{stafford2000jtb, nowak1997jv, little1999jem}, the viral load peaks around day 20 after infection. In line with \cite{chun1997nature}, there are estimated to be $\approx 10^8$ HIV-infected target cells at the viral set point. This value is about one to two orders of magnitude below peak viremia, in accordance with \cite{kinloch1995nejm, ho1996science, mcmichael2010natimm}. 
The virus can only go extinct due to stochastic effects in the beginning of infection. Lastly, the model can also realistically reproduce the simultaneous emergence of distinct viral escape mutations, as well as the generation of double escape mutants by mutation or recombination.

\subsection*{Cumulative linkage disequilibrium as a measure for interference}

\begin{figure}[h]
\begin{center}
\includegraphics[scale = 0.45]{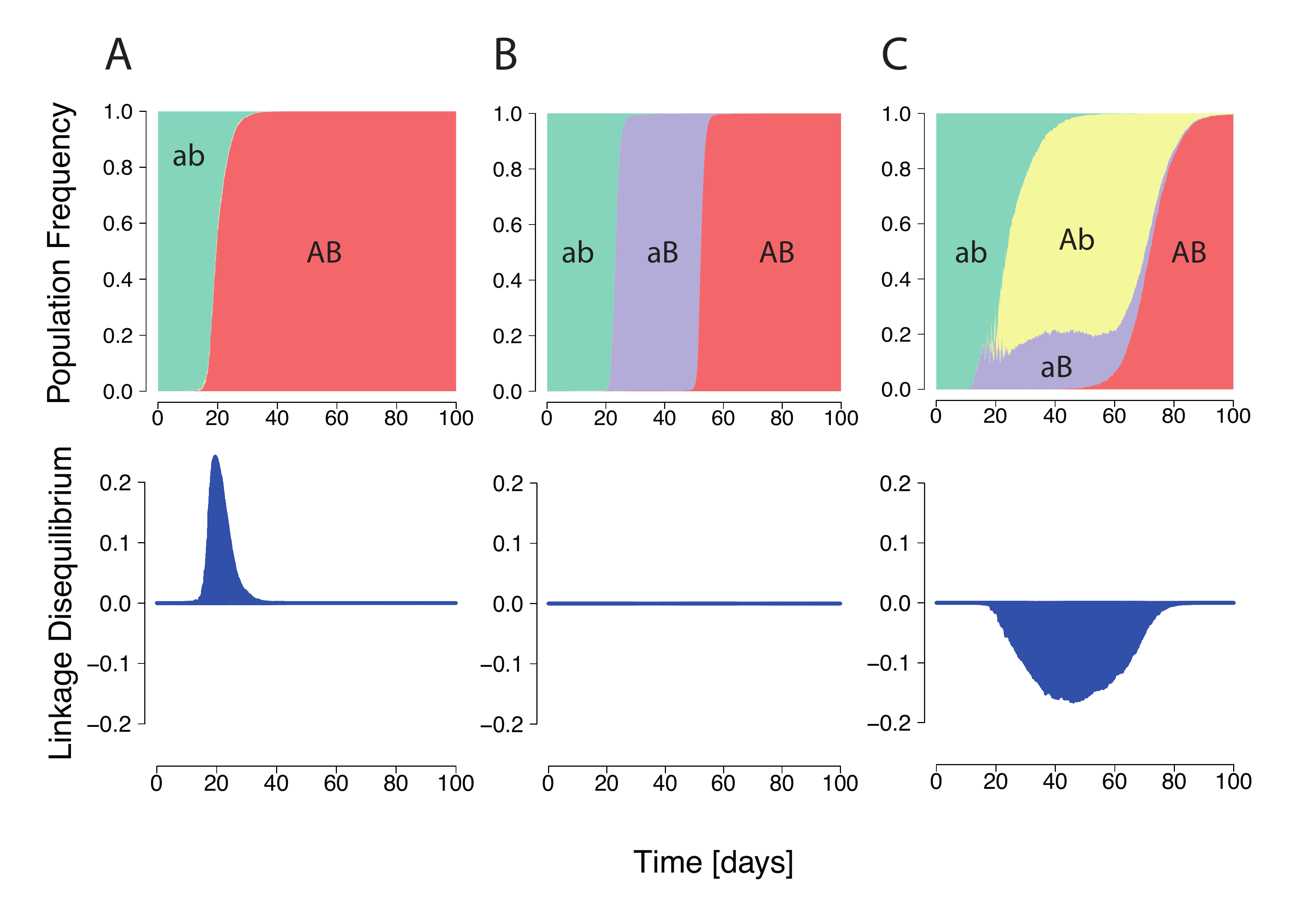} 
\end{center}
 \caption{\textbf{Three HIV dynamics scenarios and the cumulative LD.} A) The wildtype population is replaced by a double mutant. A positive LD is generated during the replacement leading to positive cumulative LD. B) Succession of wildtype, single mutant and double mutant. LD remains zero for each transition. C) Wildtype is replaced by two single mutant, which are in turn extruded by the double mutant. The longer the single mutants coexist, the more negative the cumulative LD value will be.}
 \label{fig:LD_sel}
\end{figure}

In order to quantify the \textit{expressed interference} between viral escape strains during infection, we used the population genetics measure of linkage disequilibrium (LD). In our two-epitope system (Fig.\ \ref{fig:LD_sel}), the linkage disequilibrium is $D = p_{ab}p_{AB} - p_{Ab}p_{aB}$, where $p_{ab}$ is the  frequency of the wildtype, $p_{Ab}$ and $p_{aB}$ are the frequencies of single mutants and $p_{AB}$ is the frequency of the double escape strain (where $A$ and $B$ are strongly advantageous).

%

In general, the wildtype is first replaced by single mutants, which are then outcompeted by the double mutant (Fig.\ \ref{fig:LD_sel}) \cite{tsimring1996prl, rouzine2005genetics,  desai2007genetics}. This dynamics is characterized by the duration of the intermediate phase of single mutant dominance and by the diversity in single mutants.

A quantitative measure of expressed interference of the dynamics has to behave appropriately when duration and diversity change. Firstly, the longer two single mutants coexist, the higher the value of expressed interference should be (see Fig.\ \ref{fig:LD_explan_1}). Second, the higher the diversity during the state of stasis, the higher the interference measure should be  (see Fig.\ \ref{fig:LD_explan_2}). Cumulative LD satisfies both of these conditions. The term \textit{cumulative} refers to the integral of the LD over time. This measure is also well-behaved in standard population genetics models (see Fig.\ \ref{fig:LD_explan_3}).

Negative values of the cumulative LD specifically characterize interference in a regime where selection is much stronger than recombination, and not other types of dynamics. Dynamics in which escape mutations sequentially fixate, are characterized by zero LD. 

\subsection*{Similarly strong and synchronously elicited CD$8^+$ T cell functions facilitate the appearance of interference}

In our model, we considered two CD$8^+$ T cell responses, each recognizing one of two epitopes. We investigated how differences in the strength and time delay (see \textit{Materials and Methods}) of CD$8^+$ T cell functions affect the HIV dynamics. Each CD$8^+$ T cell time course could assume two different values for its strength: $K_j \in \lbrace 1 \cdot 10^7, 7 \cdot 10^6 \rbrace$, where $j \in \lbrace 1,2 \rbrace$ denotes the order of elicitation. 
The first response was set to start at $t=0$, and the second started with a delay of 0, 5, 10, 15, 20, 25, or 30 days. Each combination of CD$8^+$ T cell function pairs was simulated 100 times. 


We consider two levels of stochasticity here: a low and a high level (see \textit{Materials and Methods}). 
The low level is displayed in our simulations at census population size of about $10^8$ at set point with parameters as in Table~\ref{table:PAR}. 
The high level of stochasticity arises when we scale the population sizes by a factor $a=10^{-4}$ (see \textit{ESM}), consistent with empirical estimates of HIV's effective population size \cite{achaz2004mbe, brown1997pnas}. 

\subsubsection*{Negative cumulative LD most pronounced at near-identical CD$8^+$ T cell responses}
\begin{figure}
\begin{center}
\includegraphics[scale = 0.5]{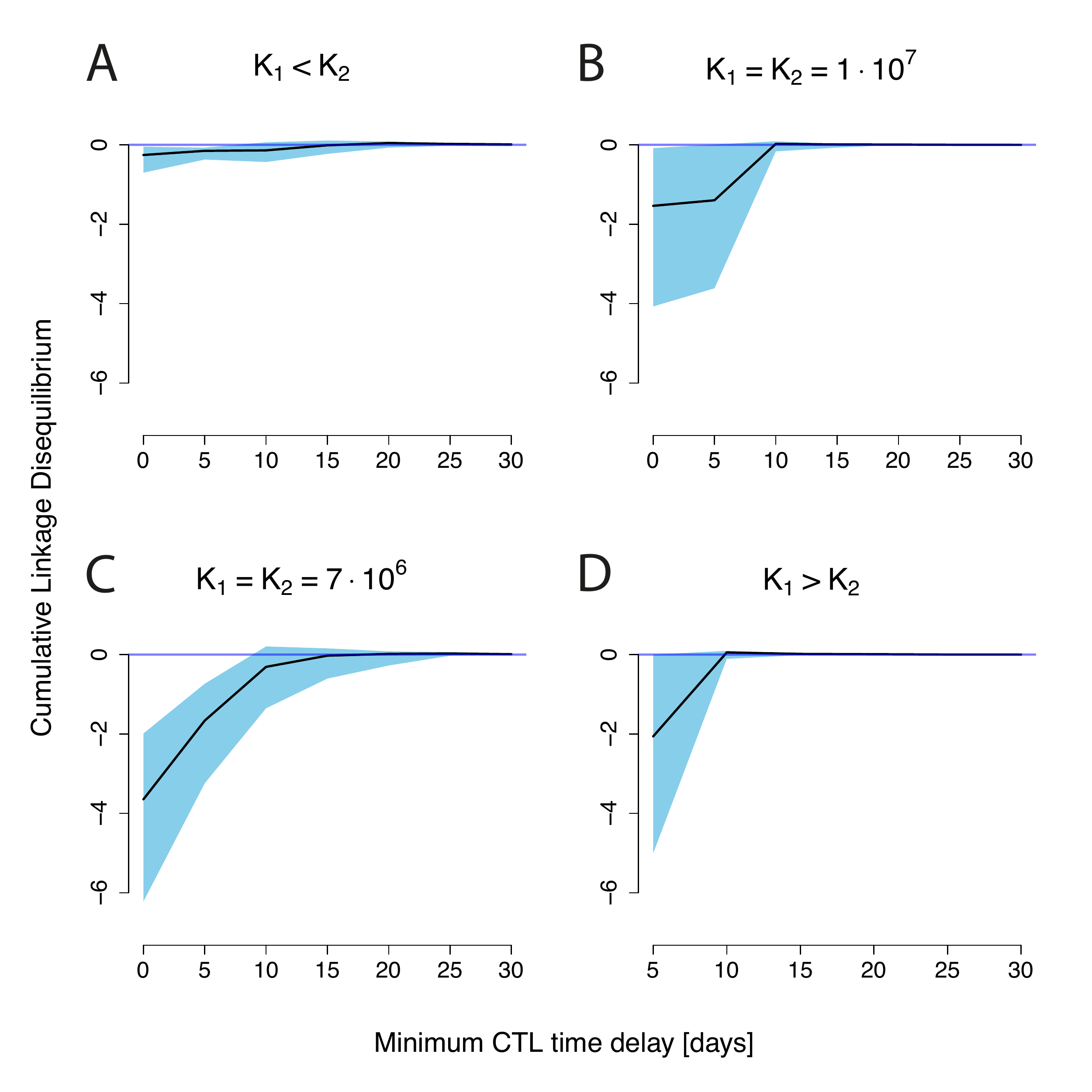} 
\end{center}
 \caption{\textbf{Cumulative LD values for simulation runs of a scaled down two-locus system ($a = 10 ^{-4}$) differing in CD$8^+$ T cell function strength and timing}. 
The x-axis denotes the time delay the second immune response has to the first. The black line is the median of 100 simulations, the upper and lower end of the blue-shaded area are the 75 and 25 percentiles of all measured simulation runs, respectively. $K_1$ and $K_2$ denote the settling values for the first and the second immune responses, respectively.  A) Little cumulative LD is generated for $K_1 < K_2$. Pronounced negative cumulative LD values are attained for nearly equally spaced CD$8^+$ T cell curves in B) and C). 
 D) shows an increase in interference at a delay of five days. }
 \label{fig:LD_s-4}
\end{figure}

To investigate under which circumstances interference between escape mutation arises we calculated the cumulative LD for different combinations of strength and timing of CD$8^+$ T cell responses, and levels of stochasticity. For the unscaled system with low stochasticity, simulations showed positive cumulative LD, indicating the immediate emergence of double escapes (see Fig. \ref{fig:LD_s1}). In contrast, figures \ref{fig:LD_s-4}A-D show negative values of cumulative LD for the down-scaled system with high stochasticity, indicating interference. Interference is particularly likely to occur when CD$8^+$ T cell responses are nearly synchronous and nearly equally strong. 


The reasons for the general pattern of negative cumulative LD are intuitively clear. First, synchronous (time delay zero), but unequally powerful CD$8^+$ T cell functions favor one of the single escape strains, leading to its fast fixation (Fig.\ \ref{fig:LD_s-4} A and D). 
This dynamics of sequential escape produces no substantial cumulative LD, and no interference. 

Second, CD$8^+$ T cell responses that are elicited far apart in time induce practically no interaction between haplotypes (Fig.\ \ref{fig:LD_s-4} B and C). The earlier immune response will select for a first escape mutant, and the second elicited immune response will select for the double escape mutant. Again, the escape dynamics is sequential, which leaves no trace in the cumulative LD.

Third, higher selective pressures (higher $K$'s) will decrease the time of emergence as well as the fixation time of double mutants. The coexistence time of single mutants will thereby be reduced. Therefore the cumulative LD, which scales roughly as the coexistence time, will be smaller compared to lower selective pressures (figures \ref{fig:LD_s-4} B and C).

There is one exception to the general pattern: when the first immune response is stronger than the second $K_1 > K_2$, and precedes it by 5 days, substantial amounts of negative cumulative LD are generated. This result arises through the complex interplay between the timing and strength of the two responses in this scenario. With a delay of 5 days, the total population of productively infected cells is at a local minimum at about the time when the action of the second response is at its peak. Due to the contracting dynamics of the CD$8^+$ T cell responses, the second response is stronger than the first at that time point, even for $K_1 > K_2$. The difference between the responses at that time point is small enough to allow for the emergence of interference. 

\subsubsection*{Escape rate decrease value}
For each simulation, we calculated the \textit{escape time} $\tau_{50}$ and \textit{escape rate} $\epsilon$ from the frequency of both escape mutations (disregarding their linkage), mimicking experimental procedures as in \cite{asquith2006inefficient,goonetilleke2009first,ganusov2011fitness,henn2012plospath} (see  \textit{ESM}). 

The estimated escape rates were in good accordance with common values for escape rates during early infection. For example, escape rates at high interference conditions (Figure \ref{fig:LD_s-4}C) were about $0.04 \hspace{0.5mm} \mathrm{day}^{-1}$ in the median, $0.02 \hspace{0.5mm} \mathrm{day}^{-1}$ for the 2.5\%-quantile and $0.17 \hspace{0.5mm} \mathrm{day}^{-1}$ for the 97.5\%-quantile. 

With the values $\tau_{50}$ and $\epsilon$ for each escape mutation, we calculated the successive escape rate decrease in each simulation. This was done by fitting a linear regression $log_{10}(\epsilon) = a + b \cdot \tau_{50}$, as in \cite{ganusov2011fitness}. The slope of the regression $b$ is termed \textit{escape rate decrease value} (ERD value). 
Negative ERD values indicate that later escapes are slower. 

For small time delays, the ERD values are between -0.01 and 0 (see Fig. \ref{fig:slope_s-4}). In \cite{ganusov2011fitness,goonetilleke2009first}, ERD values inferred from escapes with $\tau_{50}$ within the first two years, are about $-0.006$ in patient CH44, $-0.008$ in patient CH77 and $-0.01$ in patient CH58 (based on data in supplementary material of \cite{ganusov2011fitness}). 


\subsection*{Cumulative LD is associated with escape rate decrease}

To investigate the effects of interference on the ERD values we focused on those time delays shown in Fig.\ \ref{fig:LD_s-4} that showed substantial interference. 

\begin{figure}
\begin{center}
\includegraphics[scale = 0.7]{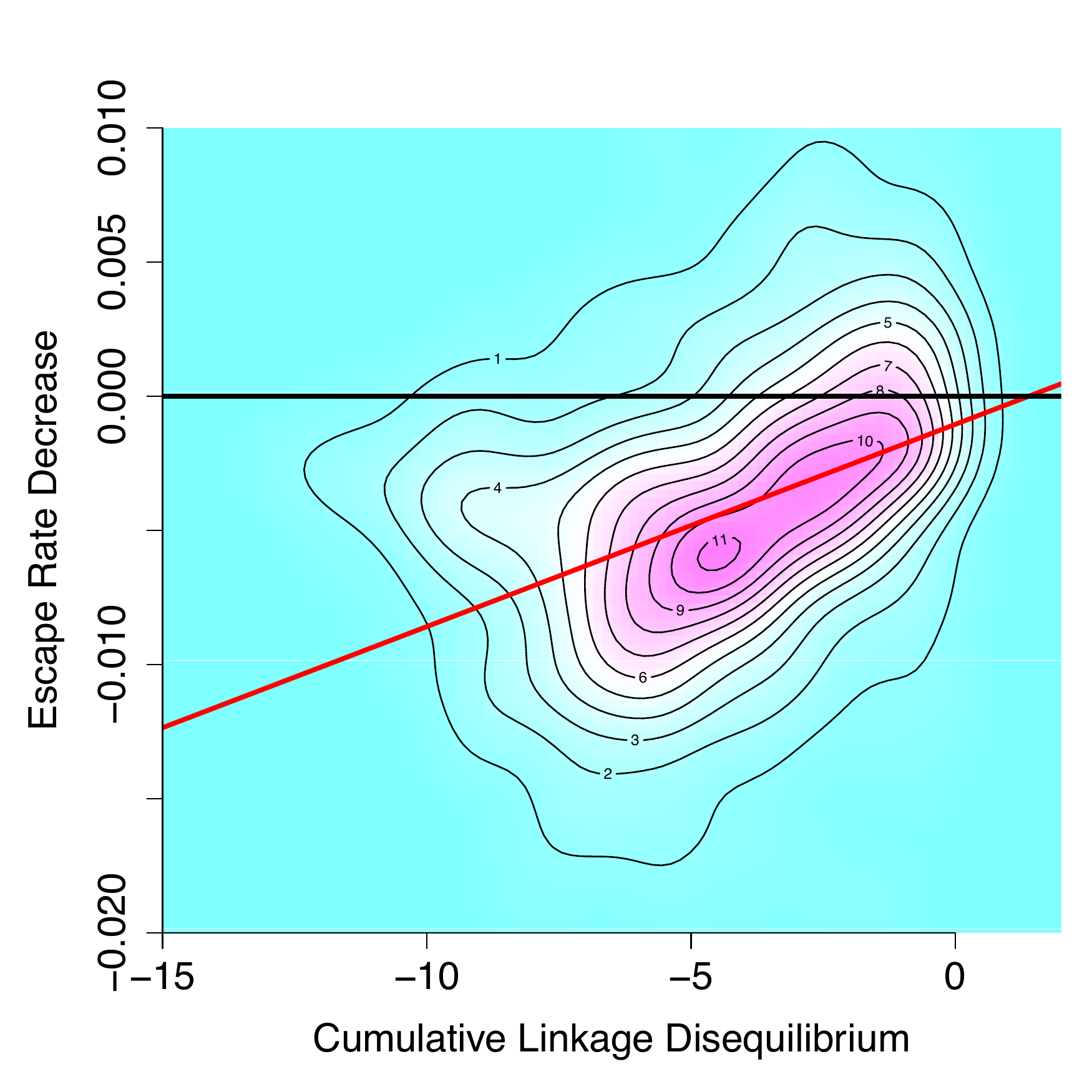} 
\end{center} 
 \caption{\textbf{Density plot of negative cumulative LD versus ERD values for 1000 simulations runs for equal CD$8^+$ T cell final values of $K_1 = K_2 = 7 \cdot 10^6$ and zero time delay between the elicitation of the CD$8^+$ T cell functions}. Positive cumulative LD values were ignored. Black line: Base line through the origin. Red line: Theil-Sen Estimator fit on data. The density distribution is compressed along a line of positive slope, indicating a positive association between interference and ERD values.}
\label{fig:Main_Result}
\end{figure}

Figure \ref{fig:Main_Result} shows the association between cumulative LD and ERD values in 1000 simulations run for CD$8^+$ T cell functions with equal final levels $K_1 = K_2 = 7 \cdot 10^6$ and no time delay. In the plot, simulations which lead to a positive cumulative LD have been removed in order to assess the effects of interference only. 
The density distribution shows a clustering of simulation results along a line of positive slope. The distribution is compressed along that line.

A Theil-Sen estimator fit to the data (red line) yields a slope of $8 \times 10^{-4}$ with confidence intervals $(6 \times 10^{-4}, 9 \times 10^{-4})$ for the 2.5 and 97.5 percentiles, respectively. This indicates an association of interference and ERD. The pattern shown in Fig.\ref{fig:Main_Result} can also be identified in the other case with identical CD$8^+$ T cell final levels ($K_1 = K_2 = 10^7$) with no time delay (see Fig.\ \ref{fig:six_results}A). 


We considered other combinations of final levels and time delays with median cumulative LD below minus one in the down-scaled system. As time delays increase, in  simulations with CD$8^+$ T cell responses of equal strength we observe the appearance of a density peak alongside the interference, centered at about zero LD and negative ERD values. This peak indicates the appearance of a different mode of escapes. These escapes are sequential and show no interference. As expected, this second mode eventually replaces the interference pattern as the time delay increases (see Fig.\ \ref{fig:six_results}).

\subsection*{A possible instance of escape rate decrease due to interference}

To investigate one possible instance of interference and its effects on escape rates, we analyzed haplotype data obtained by deep sequencing in Henn et al. \cite{henn2012plospath} of a single patient (subject 9213) infected with HIV. In this study, Blood samples were taken at days 0, 3, 59, 165, 476 and 1543 after infection was determined. Haplotypes were subsequently reconstructed from these sequence data by Pandit et al.\ \cite{pandit2013arxiv}. 

These data are of particular interest to test the effects predicted by our model because CTL responses specific to two epitopes, \textit{Nef A24-RW8} and \textit{Vif B38-WI9} where similarly strong. At day 59, these responses differed only about 10\% (see Supplementary Information in \cite{henn2012plospath}). 

\begin{figure}[h]
\begin{center}
\includegraphics[scale = 0.5]{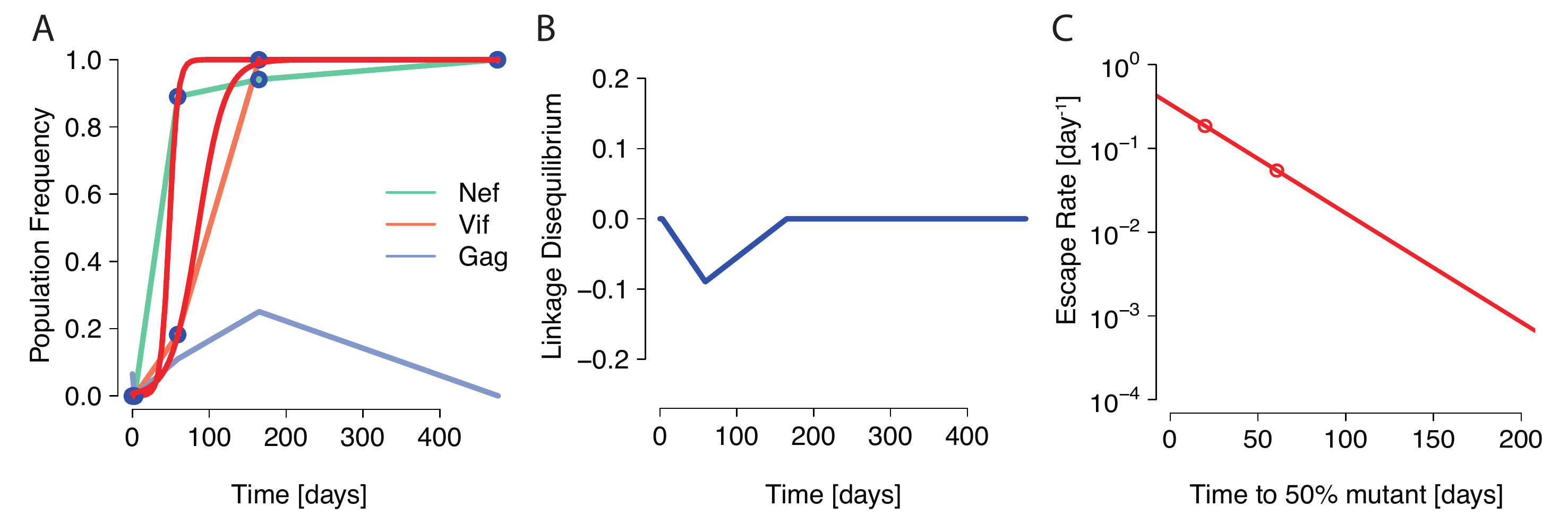} 
\end{center} 
 \caption{\textbf{A possible instance of interference affecting ERD in \cite{henn2012plospath}}. A) shows the fits of the logistic escape model to the sample points of escape mutant frequency data. B) shows the LD between \textit{Nef A24-RW8} and \textit{Vif B38-WI9} at the sample points (connected by blue line). C) shows the ERD of the escape rates of Nef and Vif. }
\label{fig:henn_analysis}
\end{figure}

To test the predictions of our model, we fitted logistic escape functions to the data and measured LD and escape rates. Figure \ref{fig:henn_analysis} shows the fitted escape curves on the mutant frequencies of \textit{Nef} and \textit{Vif}, respectively, in A). 
At day $59$ the inferred linkage disequilibrium in these data is $D = - 0.09$, as shown in B). As shown in C), ERD is clear in the escape of these two mutations.

This analysis supports the notion that interference was acting to delay the emergence of escapes and to reduce their escape rates. 



\section*{Discussion}

In this paper, the main focus was the interference of viral strains and its role in the escape dynamics. 
We identified signals of interference under very specific conditions: in systems with high levels of stochasticity and highly synchronized and comparably strong CD$8^+$ T cell responses, the generation of interference is facilitated. These conditions are often satisfied in experimental observations of HIV dynamics \cite{achaz2004mbe, brown1997pnas, antia2003jtb,murali1998immunity, deBoer2001jv}
Furthermore, when sampling at times typical for experimental studies, increasing interference decreases ERD values. Lastly, we also tested these predictions in one instance of two equally strong CTL responses elicited againts HIV in a patient \cite{henn2012plospath}. In the data a signal for interference was accompanied by ERD. 

We restricted the simulations to two-locus systems, while ERD was inferred from experimental data involving more escape variants. 
Our results are consistent with theoretical findings for two-locus systems in the context of population genetics \cite{barton1995genetics}. Intuitively, the principle that under interference the fast escape of one allele implies the slow escape of competing alleles with similar fitness should also hold in systems with more than two loci. 
We thus hypothesize that the pattern of ERD is preserved in HIV dynamics models with more than two epitopes showing interference. 

Furthermore, we ignored potential fitness costs of escape mutations in our simulations. 
They can be safely neglected if they are compensated at faster rates than the fixation times of beneficial mutants. Only if fitness costs differ substantially between escape mutants do we expect that their explicit consideration will alter the role of interference.

These results have to be interpreted in the larger context of estimating the selective pressures that immune responses exert on the virus population. These selection pressures are often inferred from the growth advantage of mutants that escaped the immune response --- the rate of escape. Estimating these rates of escape using models that neglect the complex genetical interactions between escape strains has revealed the pattern of ERD central to our study. Our study shows that caution is warranted when drawing conclusions from this pattern about the selection pressures at work.

Very recently, Pandit et.\ al.\ identified a clear instance of clonal interference in HIV between mutations within the same epitope 
as well as between epitopes, \cite{pandit2013arxiv}. 
The relevance of interference during early HIV is further supported by "epitope shattering", where a founder strain can diversify into an array of strains with distinct escape mutations at the same epitope \cite{boutwell2010jid}. Leviyang studied such mutational pathways in data presented in \cite{fischer2010plosone}, focusing on competition between intra-epitope escape mutations \cite{leviyang2013genetics}. O'Connor also reported the coexistence of escape mutations within the same epitope in SIV-infected Mauritian cynomolgus macaques \cite{o2012conditional}. 


In contrast to our investigations, other studies expect interference effects to be negligible. da Silva modeled early HIV infection with a Wright-Fisher process incorporating weakening CD$8^+$ T cell responses \cite{da2012genetics}. 
He concluded that due to the transmission bottleneck the effective population size of HIV should remain at low levels ($N_e \approx 10^2$) throughout early infection, thus preventing interference. 

Kessinger et al.\ \cite{kessinger2013fii} estimated escape rates of the HIV data presented in Goonetilleke et al.\ \cite{goonetilleke2009first} by employing multi-epitope models of HIV. To do this, they imposed a scheme of sequential escapes on their model. 
With that scheme, 
the escape rate estimates were substantially higher than in other studies.

Very recently, Ganusov et al.\ presented stochastic simulations of a multi-epitope model of HIV infection with recombination \cite{ganusov2013jsm}. In that paper, the bias interference effects introduce in escape rate estimates is also discussed. Escape rate estimates are heavily underestimated at low sample sizes ($\approx 20$ samples), but improve at sample sizes of about 200. They also find that in stochastic simulations, escapes are delayed compared to deterministic escapes, especially for low recombination rates. These theoretical results strongly support the notion that current estimation methods might be inappropriate tools for escape rate inference under interference regimes. 

Our study makes a few testable predictions. Whether interference is involved in the generation of ERD can be assessed by measuring the linkage disequilibrium between HIV haplotypes over time. This requires sequencing that either retains linkage information, or the reconstruction of haplotypes using bioinformatic methods \cite{zagordi2011bmc, prosperi2012bioinf}. Sustained negative linkage disequilibria would be indicative of interference. In this case, interference needs to be corrected before escape rates can be related to selection pressures. After this correction, the estimates of the selection pressures should be higher than previously estimated by means of logistic curve fitting.  
Furthermore, interference should be enhanced in mutations of epitopes in close proximity in the genome, especially mutations within the same epitopes.


\section*{Acknowledgments}
The authors gratefully acknowledge the funding of the Swiss National Science Foundation (grant number 315230-130855 for V.G. and R.R.R, P1EZP3\_148648 for V.G.). 




\newpage



\setcounter{figure}{0}
\makeatletter 
\renewcommand{\thefigure}{S\@arabic\c@figure}
\makeatother

%
%
%

\begin{figure}[h]
\begin{center}
\end{center}
\caption{ } 
 \label{fig:LD_explan_1}
\end{figure}

\begin{figure}[h]
\begin{center}
\end{center}
\caption{ }
 \label{fig:LD_explan_2}
\end{figure}

\begin{figure}[h]
\begin{center}
\end{center}
\caption{ }
 \label{fig:LD_explan_3}
\end{figure}

\begin{figure}[h]
\begin{center}
\end{center} 
\caption{ }
\label{fig:recombination}
\end{figure}

\begin{figure}[h]
\begin{center}
\end{center}
\caption{ }
 \label{fig:LD_s1}
\end{figure}

\begin{figure}[h]
\begin{center}
\end{center}
\caption{ } 
 \label{fig:slope_s-4}
\end{figure}

\begin{figure}[h]
\begin{center}
 \end{center}
 \caption{ }
\label{fig:six_results}
\end{figure}

\newpage


\end{document}